\documentstyle[preprint,aps,epsfig]{revtex}
\begin{document}
\tightenlines
\draft
\title{Limits on Anomalous $WW\gamma$ and $WWZ$ Couplings
 }
\maketitle
\begin{center}
%
B.~Abbott,$^{31}$
M.~Abolins,$^{27}$
B.~S.~Acharya,$^{46}$
I.~Adam,$^{12}$
D.~L.~Adams,$^{40}$
M.~Adams,$^{17}$
S.~Ahn,$^{14}$
H.~Aihara,$^{23}$
G.~A.~Alves,$^{10}$
N.~Amos,$^{26}$
% ...
\\
\vskip 0.50cm
\centerline{(D\O\ Collaboration)}
\vskip 0.50cm
\sl{
\centerline{$^{1}$Universidad de los Andes, Bogot\'{a}, Colombia}
\centerline{$^{2}$University of Arizona, Tucson, Arizona 85721}
\centerline{$^{3}$Boston University, Boston, Massachusetts 02215}
\centerline{$^{4}$Brookhaven National Laboratory, Upton, New York 11973}
\centerline{$^{5}$Brown University, Providence, Rhode Island 02912}
\centerline{$^{6}$Universidad de Buenos Aires, Buenos Aires, Argentina}
\centerline{$^{7}$University of California, Davis, California 95616}
\centerline{$^{8}$University of California, Irvine, California 92697}
\centerline{$^{9}$University of California, Riverside, California 92521}
\centerline{$^{10}$LAFEX, Centro Brasileiro de Pesquisas F{\'\i}sicas,
                  Rio de Janeiro, Brazil}
\centerline{$^{11}$CINVESTAV, Mexico City, Mexico}
\centerline{$^{12}$Columbia University, New York, New York 10027}
\centerline{$^{13}$Delhi University, Delhi, India 110007}
\centerline{$^{14}$Fermi National Accelerator Laboratory, Batavia,
                   Illinois 60510}
\centerline{$^{15}$Florida State University, Tallahassee, Florida 32306}
\centerline{$^{16}$University of Hawaii, Honolulu, Hawaii 96822}
\centerline{$^{17}$University of Illinois at Chicago, Chicago,
                   Illinois 60607}
\centerline{$^{18}$Indiana University, Bloomington, Indiana 47405}
\centerline{$^{19}$Iowa State University, Ames, Iowa 50011}
\centerline{$^{20}$Korea University, Seoul, Korea}
\centerline{$^{21}$Kyungsung University, Pusan, Korea}
\centerline{$^{22}$Institute of Nuclear Physics, Krak\'ow, Poland}
\centerline{$^{23}$Lawrence Berkeley National Laboratory and University of
                   California, Berkeley, California 94720}
\centerline{$^{24}$Louisiana Tech University, Ruston, Louisiana 71272}
\centerline{$^{25}$University of Maryland, College Park, Maryland 20742}
\centerline{$^{26}$University of Michigan, Ann Arbor, Michigan 48109}
\centerline{$^{27}$Michigan State University, East Lansing, Michigan 48824}
\centerline{$^{28}$Institute for Theoretical and Experimental Physics,
                   Moscow, Russia}
\centerline{$^{29}$Moscow State University, Moscow, Russia}
\centerline{$^{30}$University of Nebraska, Lincoln, Nebraska 68588}
\centerline{$^{31}$New York University, New York, New York 10003}
\centerline{$^{32}$Northeastern University, Boston, Massachusetts 02115}
\centerline{$^{33}$Northern Illinois University, DeKalb, Illinois 60115}
\centerline{$^{34}$Northwestern University, Evanston, Illinois 60208}
\centerline{$^{35}$University of Notre Dame, Notre Dame, Indiana 46556}
\centerline{$^{36}$University of Oklahoma, Norman, Oklahoma 73019}
\centerline{$^{37}$University of Panjab, Chandigarh 16-00-14, India}
\centerline{$^{38}$Institute for High Energy Physics, Protvino 142284,
                   Russia}
\centerline{$^{39}$Purdue University, West Lafayette, Indiana 47907}
\centerline{$^{40}$Rice University, Houston, Texas 77005}
\centerline{$^{41}$Universidade do Estado do Rio de Janeiro, Brazil}
\centerline{$^{42}$University of Rochester, Rochester, New York 14627}
\centerline{$^{43}$CEA, DAPNIA/Service de Physique des Particules,
                   CE-SACLAY, Gif-sur-Yvette, France}
\centerline{$^{44}$Seoul National University, Seoul, Korea}
\centerline{$^{45}$State University of New York, Stony Brook,
                   New York 11794}
\centerline{$^{46}$Tata Institute of Fundamental Research,
                   Colaba, Mumbai 400005, India}
\centerline{$^{47}$University of Texas, Arlington, Texas 76019}
\centerline{$^{48}$Texas A\&M University, College Station, Texas 77843}
}
\end{center}

\date{\today}

\begin{abstract}
Limits on the anomalous $WW\gamma$ and $WWZ$ couplings are presented from
a simultaneous fit to the data samples of three gauge boson pair
final states in $p\bar{p}$ collisions at $\sqrt{s}=1.8$ TeV:
$W\gamma$ production with the $W$ boson decaying to $e\nu$ or $\mu\nu$,
$W$ boson pair production with both of the $W$ bosons decaying to $e\nu$ or
$\mu\nu$, and $WW$ or $WZ$ production with one $W$ boson decaying to $e\nu$ and
the other $W$ boson or the $Z$ boson decaying to two jets.
Assuming identical $WW\gamma$ and $WWZ$ couplings,
$95 \%$ C.L. limits on the anomalous
couplings of $-0.30<\Delta\kappa<0.43 ~(\lambda = 0)$
and $-0.20<\lambda<0.20 ~(\Delta\kappa = 0)$ are obtained
using a form factor scale $\Lambda = 2.0$ TeV.
Limits found under other assumptions on the relationship between the
$WW\gamma$ and $WWZ$ couplings are also presented.
\end{abstract}



Gauge boson self-interactions are a direct consequence of
the non-Abelian $SU(2)\times U(1)$ gauge symmetry of the
standard model (SM) and
are a necessary element to construct unitary and renormalizable
theories involving massive gauge bosons~\cite{theory}.
The values of trilinear gauge boson couplings
are fully determined in the SM by the gauge structure.
The precise determination of the couplings constitutes one of few
remaining tests of the SM; any deviation from the SM
values would indicate the presence of new physics.
Phenomenological bounds on the trilinear gauge boson couplings
have been obtained from the precisely measured quantities, such as
$(g-2)_{\mu}$, the $b\rightarrow s\gamma$ decay rate,
the $Z\rightarrow b{\bar b}$ rate
and oblique corrections~\cite{indirect}.
These bounds are obtained with many assumptions imposed on
the couplings.
The trilinear gauge boson couplings can be measured directly with
fewer assumptions by studying gauge boson pair production processes.
Direct measurements of the couplings have been reported by the
UA2~\cite{UA2}, CDF~\cite{CDF,CDFWWWZ}, D{\O}~\cite{D0Wg,D0WW,D0enjj},
and LEP~\cite{LEP} collaborations.
Hadron collider experiments have established the electroweak
coupling of the $W$ boson to the photon~\cite{D0Wg} and the existence of the
coupling between the $W$ boson and the $Z$ boson~\cite{CDFWWWZ,D0enjj} by placing
constraints on anomalous $WW\gamma$ and $WWZ$ couplings.

The $WW\gamma$ and $WWZ$ vertices are described by a
general effective Lagrangian
with two overall coupling constants, $g_{WW\gamma} = -e$ and
$g_{WWZ} = -e \cdot \cot \theta_{W}$ (where $e$ is the $W^+$ charge and
$\theta_{W}$ is the weak mixing angle), and
six dimensionless coupling parameters,
$g_{1}^{V}$, $\kappa_V$, and $\lambda_V$ ($V = \gamma$ or $Z$),
after imposing {\it C}, {\it P}, and {\it CP} invariance~\cite{Lagrangian}.
Electromagnetic gauge invariance requires that $g_{1}^{\gamma} = 1$, which
we assume throughout this paper.
The effective Lagrangian becomes that of the SM when
$g_1^{\gamma} = g_1^Z = 1$ $(\Delta g_1^V \equiv g_1^V - 1 = 0)$,
$\kappa_{V} = 1$ $(\Delta\kappa_{V} \equiv \kappa_V - 1 = 0)$, and
$\lambda_V = 0$.
Limits on these couplings are usually obtained under the assumption that the
$WW\gamma$ and $WWZ$ couplings are equal
($g_1^{\gamma} = g_1^Z = 1$, $\Delta\kappa_{\gamma} = \Delta\kappa_Z$, and
$\lambda_{\gamma} = \lambda_Z$).

A different set of parameters, motivated by $SU(2)\times U(1)$ gauge
invariance, has been used by the LEP collaborations~\cite{alpha}.
This set consists of three independent couplings
$\alpha_{B\phi}$, $\alpha_{W\phi}$ and $\alpha_W$:
$\alpha_{B\phi}\equiv \Delta\kappa_{\gamma} - \Delta g_1^Z \cos^{2}\theta_{W}$,
$\alpha_{W\phi}\equiv\Delta g_1^Z \cos^{2}\theta_{W}$ and
$\alpha_{W}\equiv\lambda_{\gamma}$.
The remaining $WWZ$ coupling parameters $\lambda_Z$ and $\Delta\kappa_Z$
are determined by the relations $\lambda_Z = \lambda_{\gamma}$ and
$\Delta\kappa_Z = -\Delta\kappa_{\gamma}\tan^{2}\theta_{W} + \Delta g_1^Z$.
The HISZ relations~\cite{HISZ}
which have been used by the D{\O} and CDF collaborations are
also based on this set with the additional constraint
$\alpha_{B\phi} = \alpha_{W\phi}$.

Non-SM couplings give rise to a large increase in the cross section
of gauge boson pair production processes at high energies.
To avoid violation of unitarity,
the anomalous couplings are modified by form factors with a scale $\Lambda$
({\it e.g.} $\lambda_{V}(\hat{s}) =
\lambda_{V} / ( 1 + \hat{s}/\Lambda^{2})^{2}$),
which is related to the scale of new physics.

The D{\O} collaboration has previously reported
limits on anomalous $WW\gamma$ and $WWZ$ couplings from
the data samples of three gauge boson pair
final states:
$W\gamma$ production with the $W$ boson decaying to
$e\nu$ or $\mu\nu$~\cite{D0Wg},
$W$ boson pair production with both of the $W$ bosons decaying to $e\nu$ or
$\mu\nu$~\cite{D0WW},
and $WW$ or $WZ$ production with one $W$ boson decaying to $e\nu$ and
the other $W$ boson or the $Z$ boson decaying to two jets~\cite{D0enjj}.
The data samples  correspond to an integrated luminosity of
approximately 100 pb$^{-1}$ collected with the  D\O\ detector
during the 1992--93 and 1993--1995 Tevatron collider runs at Fermilab.
This report is a culmination of these studies and presents the most stringent
limits available on anomalous $WW\gamma$ and $WWZ$ couplings by performing
a simultaneous fit to the data
samples of the above three final states.
Limits are also set on the $\alpha$ parameters, enabling a direct comparison of
our results with those of LEP experiments.

The D{\O} detector and data collection system are described
elsewhere~\cite{D0Detector}.
Limits on the anomalous couplings are obtained by a maximum likelihood fit to
the transverse energy ($E_T$) spectrum of a final state gauge boson or to the
$E_T$ spectra of the decay leptons from the gauge boson pair.
Since the predicted relative increase in the gauge boson pair production cross section
with anomalous couplings is greater at higher gauge boson $E_T$,
fits to the $E_T$ spectra provide significantly tighter constraints on
anomalous couplings than those from the measurement of the cross section alone.
The individual analyses have been described in detail
previously~\cite{D0PRD}.
This paper reports only on the simultaneous fit to the three data sets.

In this analysis, as in the previous reports, a binned maximum likelihood fit
is performed to the candidate events.
The probability $P_i$ for observing $N_i$ events in a given bin of a
kinematical variable is
$P_i = e^{-(b_i + n_i)}\frac{(b_i + n_i)^{N_i}}{N_{i}!}$,
where $b_i$ is the estimated background,
$n_i ( = {\cal L}\epsilon\sigma_{i}(\lambda,\Delta\kappa))$
is the expected signal,
${\cal L}$ is the integrated luminosity,
$\epsilon$ is the detection efficiency, and $\sigma_i$ is the
theoretical cross section which is a function of anomalous couplings, $\lambda$
and $\Delta\kappa$.
The joint probability $P$ for all the kinematical bins that are fitted is
$P = \prod^{N_{bin}}_{i=1} P_i$.
Since the variables $b_i, {\cal L}$, $\epsilon$ and the normalization of the
predicted theoretical cross section are estimated
quantities with some uncertainty
and do not depend on $\lambda$ and $\Delta\kappa$,
we assign Gaussian prior distributions and integrate over the possible ranges;
$P'\propto\int {\cal G}_{f_n} df_n \int {\cal G}_{f_b} df_b
\prod^{N_{bin}}_{i=1}
\frac{e^{-(f_{n}n_{i} + f_{b}b_{i})}(f_{n}n_{i} + f_{b}b_{i})^{N_i}}
{N_{i}!}$,
where ${\cal G}_{f_b}$ and ${\cal G}_{f_n}$ are Gaussian functions with
standard deviation $\sigma_b$ and $\sigma_n$ for the background and the signal,
respectively.
For convenience, the log-likelihood, $L = - \log P'$, is used.
When the simultaneous fit is performed on the three data sets, correlations
between $\sigma_b$ and $\sigma_n$ for different final states are carefully
taken into account.

In Table~\ref{table:Wg_limits}, the $95 \%$ C.L. limits on anomalous
$WW\gamma$ couplings from the $W\gamma$
analysis are listed.
The $W\gamma$ candidate events are selected by requiring an isolated high
$E_T$ electron or muon, large missing transverse energy
(\hbox{$\rlap{\kern0.25em/}E_T$}) and an isolated high $E_T$
photon.
The limits are obtained from a binned maximum likelihood fit to the $E_T$
spectrum of the photons.
In this process, only $\lambda_{\gamma}$ and $\Delta\kappa_{\gamma}$ couplings
are involved.

In Table~\ref{table:dileptons_limits}, the $95 \%$ C.L. limits from the
$WW\rightarrow$ dilepton analysis are listed.
The $WW\rightarrow$ dilepton candidate events are selected by requiring
two high $E_T$ leptons ($ee$, $e\mu$, or $\mu\mu$) and large
\hbox{$\rlap{\kern0.25em/}E_T$}.
The limits are obtained from a maximum likelihood fit to the number of observed
candidate events in two-dimensional $E_T$ bins of the decay leptons
from the $W$ boson pair.

In Table~\ref{table:enjj_limits}, the $95 \%$ C.L. limits from
the $WW/WZ\rightarrow e\nu jj$ analysis are listed.
The $WW/WZ\rightarrow e\nu jj$ candidate events are selected by requiring
an isolated high $E_T$ electron, large \hbox{$\rlap{\kern0.25em/}E_T$},
and two high $E_T$ jets.
The invariant mass of the two jet system must be consistent with that of
the $W$ or $Z$ boson.
Limits are obtained from a binned maximum likelihood fit
to the $E_T$ spectrum
of the $W$ boson calculated from the electron $E_T$ and
$\hbox{$\rlap{\kern0.25em/}E_T$}$,
using four sets of relationships
between the $WW\gamma$ and $WWZ$ couplings:
(i) $\Delta\kappa \equiv \Delta\kappa_\gamma = \Delta\kappa_Z,
\lambda \equiv \lambda_\gamma = \lambda_Z$,
(ii) HISZ relations, (iii) varying the $WWZ$ couplings
while the $WW\gamma$ couplings are fixed to the SM values, and (iv) varying
the $WW\gamma$ couplings while the $WWZ$ couplings are fixed to the SM values.
Two values of $\Lambda$, 1.5 and 2.0 TeV, are used.

\begin{figure}[h]
\centerline{\epsfig{figure=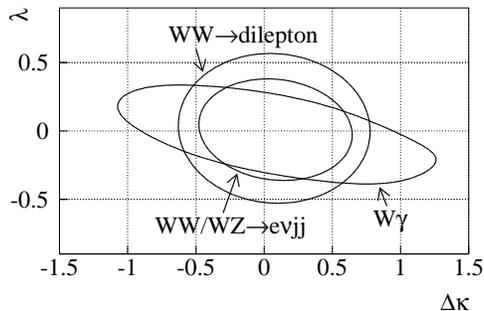,width=2.8in}}
 \caption{Contour limits on anomalous couplings for $\Lambda = 1.5$ TeV.
 For the $WW\rightarrow$ dilepton and $WW/WZ\rightarrow e\nu jj$
 contour limits, the $WW\gamma$ and $WWZ$ couplings are assumed to be equal.
 The contours plotted in Figs.~\ref{fig:three}--\ref{fig:lep_sets} are
 accurate to $\pm 0.02$ due to MC statistics.}
 \label{fig:three}
\end{figure}

Tables~\ref{table:Wg_limits}--\ref{table:enjj_limits} are reproduced from the
previous reports.
Figure~\ref{fig:three} contains the $95\%$ C.L. one-degree of freedom
exclusion contours~\cite{CL} from the $W\gamma$,
$WW\rightarrow$ dilepton, and $WW/WZ\rightarrow e\nu jj$ analyses.
The contours that represent the unitarity constraint~\cite{Unitarity}
for individual processes are omitted in Fig.~\ref{fig:three}.
 
\begin{figure}[h]
\centerline{\hbox{
\epsfig{figure=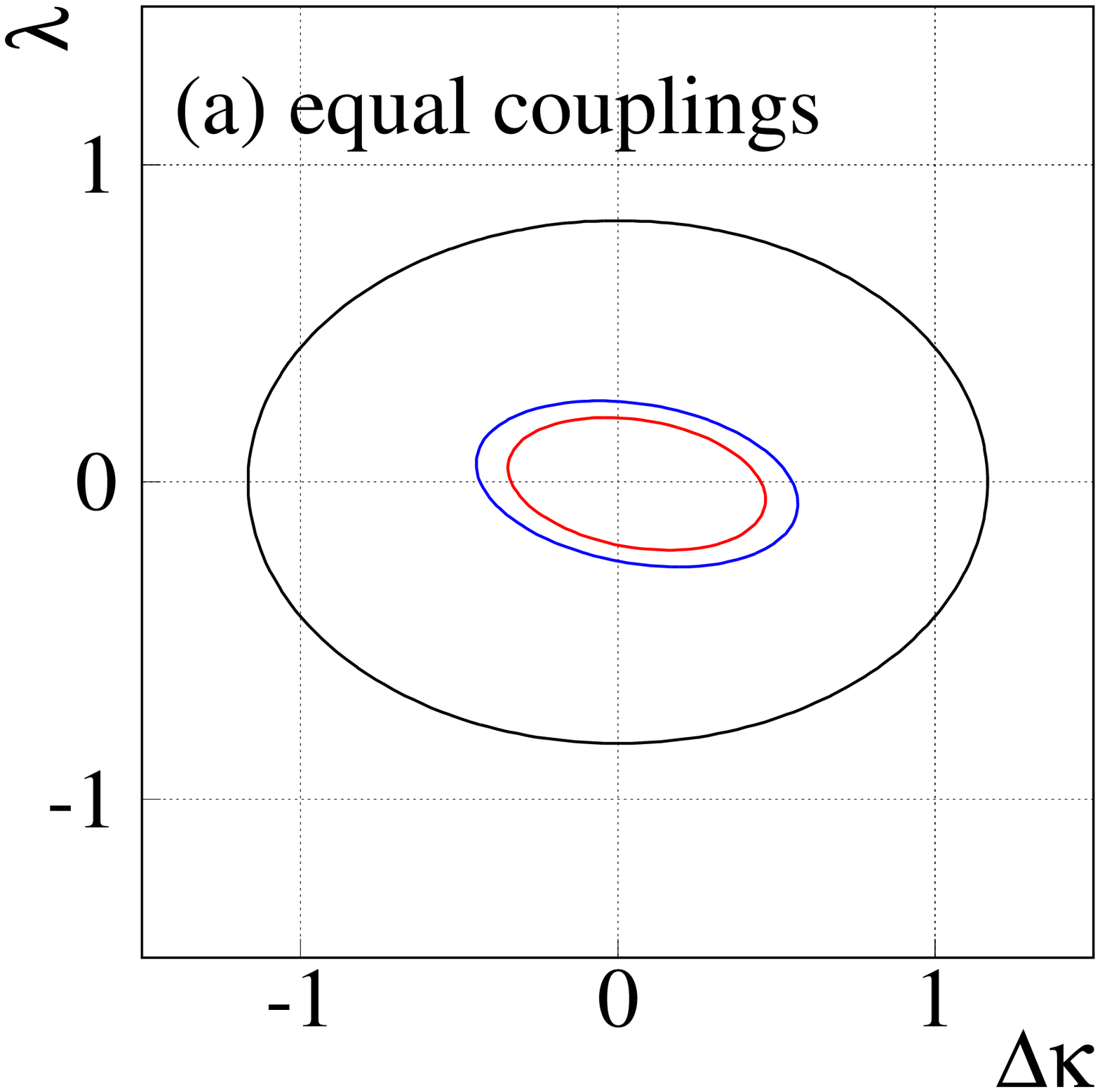,width=1.8in}
\epsfig{figure=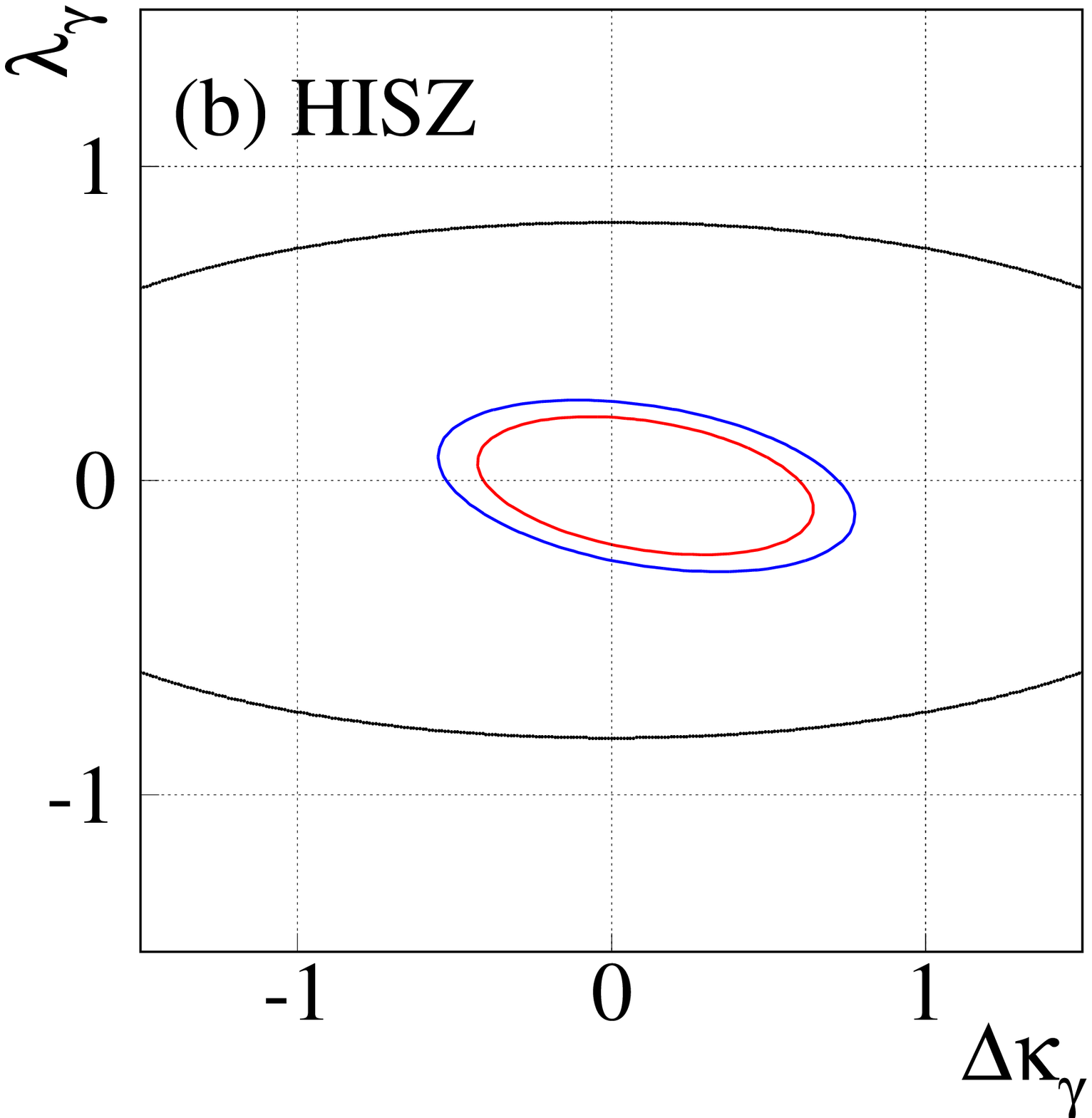,width=1.8in}}}
\centerline{\hbox{
\epsfig{figure=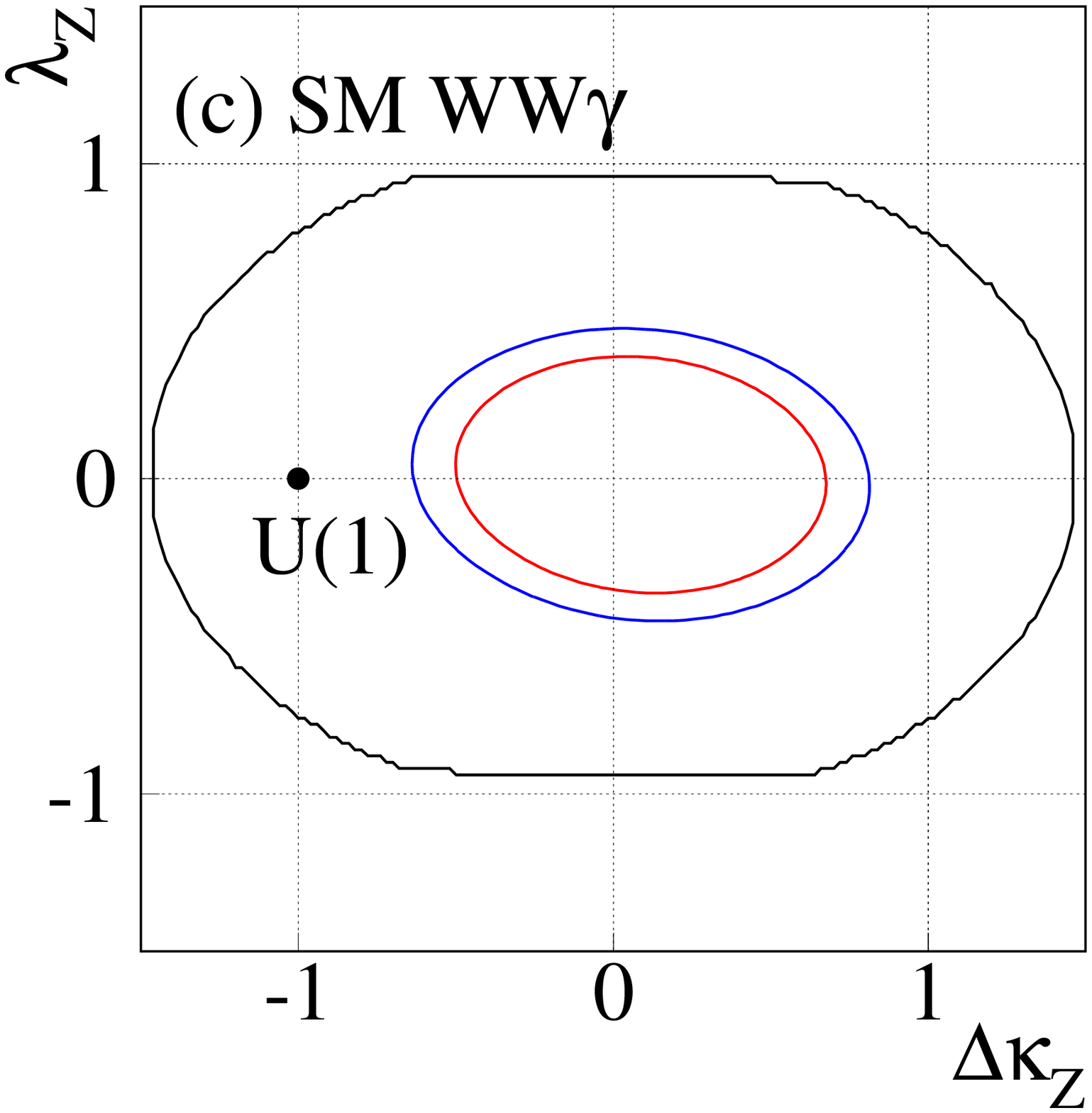,width=1.8in}
\epsfig{figure=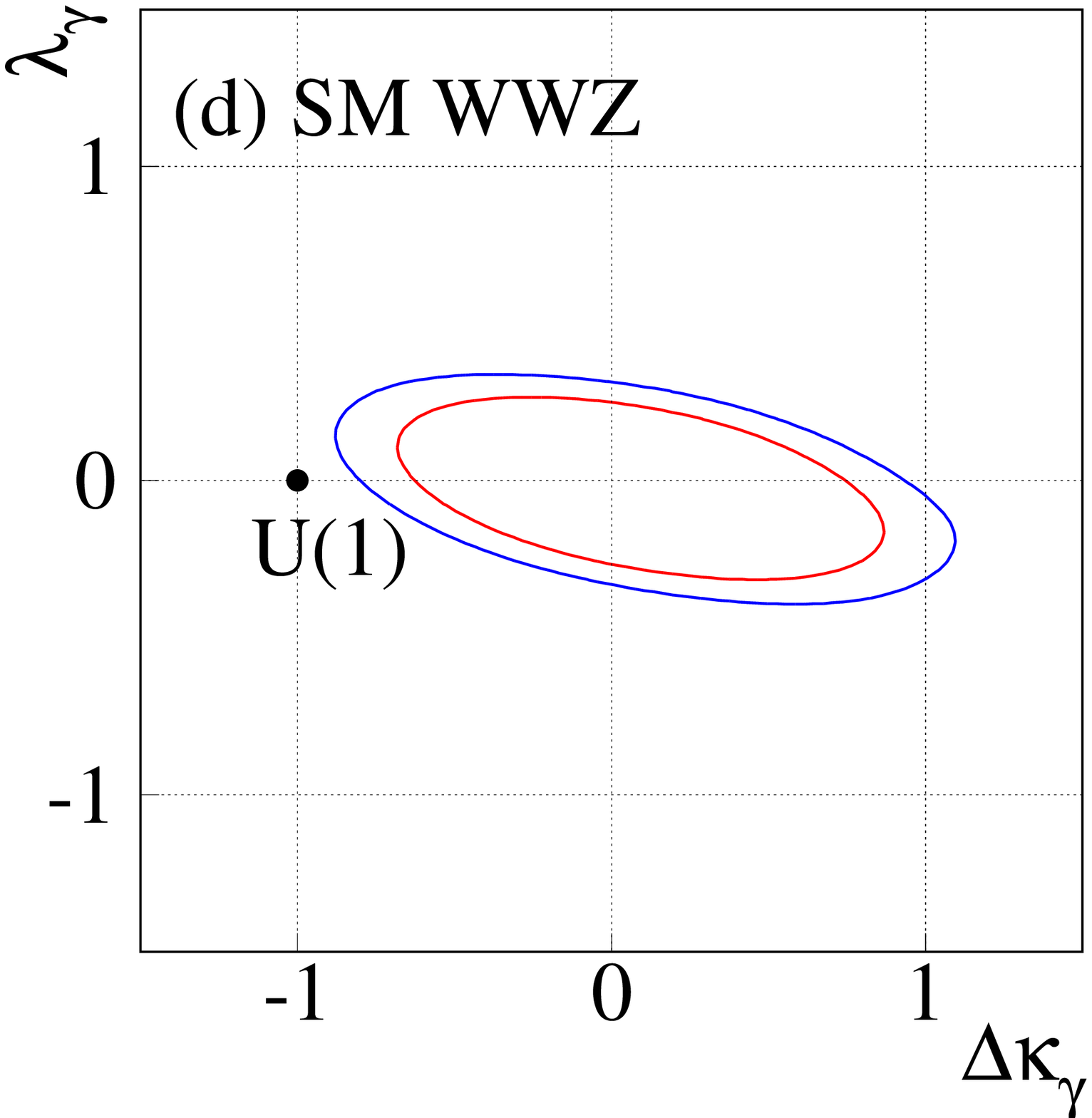,width=1.8in}}}
 \caption{Contour limits on anomalous couplings
 from a simultaneous fit to the data sets of $W\gamma$, $WW\rightarrow$
 dilepton, and $WW/WZ\rightarrow e\nu jj$ final states
 for $\Lambda = 1.5$ TeV:
 (a) $\Delta\kappa \equiv \Delta\kappa_\gamma = \Delta\kappa_Z,
 \lambda \equiv \lambda_\gamma = \lambda_Z$; (b) HISZ relations;
 (c) SM $WW\gamma$ couplings; and (d) SM $WWZ$ couplings.
 (a), (c), and (d) assume that $\Delta g_1^Z=0$.
 The innermost and middle curves are
 $95 \%$ C.L. one- and two-degree of freedom exclusion contours, respectively.
 The outermost curve is the constraint from the unitarity
 condition. In (d), the unitarity contour is located outside of the
 boundary of the plot.}
 \label{fig:combined}
\end{figure}

In Table~\ref{table:combined_limits}, the $95 \%$ C.L. limits from
a simultaneous fit to the three data sets are presented.
The common uncertainties, those on the integrated luminosity ($5.3\%$) and
the theoretical cross section of the gauge boson pair production ($7\%$),
are factored out and included only once in the integration.
Correlations in the uncertainties on the electron and muon selection efficiencies
between the data sets of the 1992--1993 and 1993--1995 runs are properly
taken into account for individual final states.
Correlations in the uncertainties on the electron and muon selection efficiencies
between different final states are
ignored, since the uncertainties themselves are small and have practically
no effect on the limits.
Correlations in the uncertainties on the background estimates between the
data sets of the 1992--1993 and 1993--1995 runs are properly
taken into account for individual final states.
The uncertainties on the background estimates between
different final states are assumed to be uncorrelated,
since the dominant sources of
uncertainties are different for the three final states.
Figure~\ref{fig:combined}(a) shows the contour limits when the $WW\gamma$ and
$WWZ$ couplings are assumed to be equal.
Figure~\ref{fig:combined}(b) shows the contour limits assuming HISZ relations.
In Fig.~\ref{fig:combined}(c), the contour limits on anomalous $WWZ$
couplings  are shown assuming the SM $WW\gamma$ couplings.
The $U(1)$ point ($\kappa_Z=0$, $\lambda_Z=0$ and
$g_1^Z=0$) indicated in the figure,
which implies that there is no coupling between the $W$
boson and the $Z$ boson, is excluded at the 99.99\% C.L.
In Fig. \ref{fig:combined}(d), the contour limits on anomalous $WW\gamma$
couplings are shown assuming the SM $WWZ$ couplings.
The $U(1)$ point ($\kappa_\gamma=0$, and $\lambda_\gamma=0$)
indicated in the figure,
which implies that the $W$ boson couples to the photon with the
electromagnetic interactions only, is excluded at the 99.7\% C.L.
The innermost and middle curves are
$95 \%$ C.L. one- and two-degree of freedom exclusion contours,
respectively~\cite{CL}.
The outermost curve is the constraint from the unitarity
condition with $\Lambda=1.5$ TeV.

\begin{figure}[h]
\centerline{\hbox{
\epsfig{figure=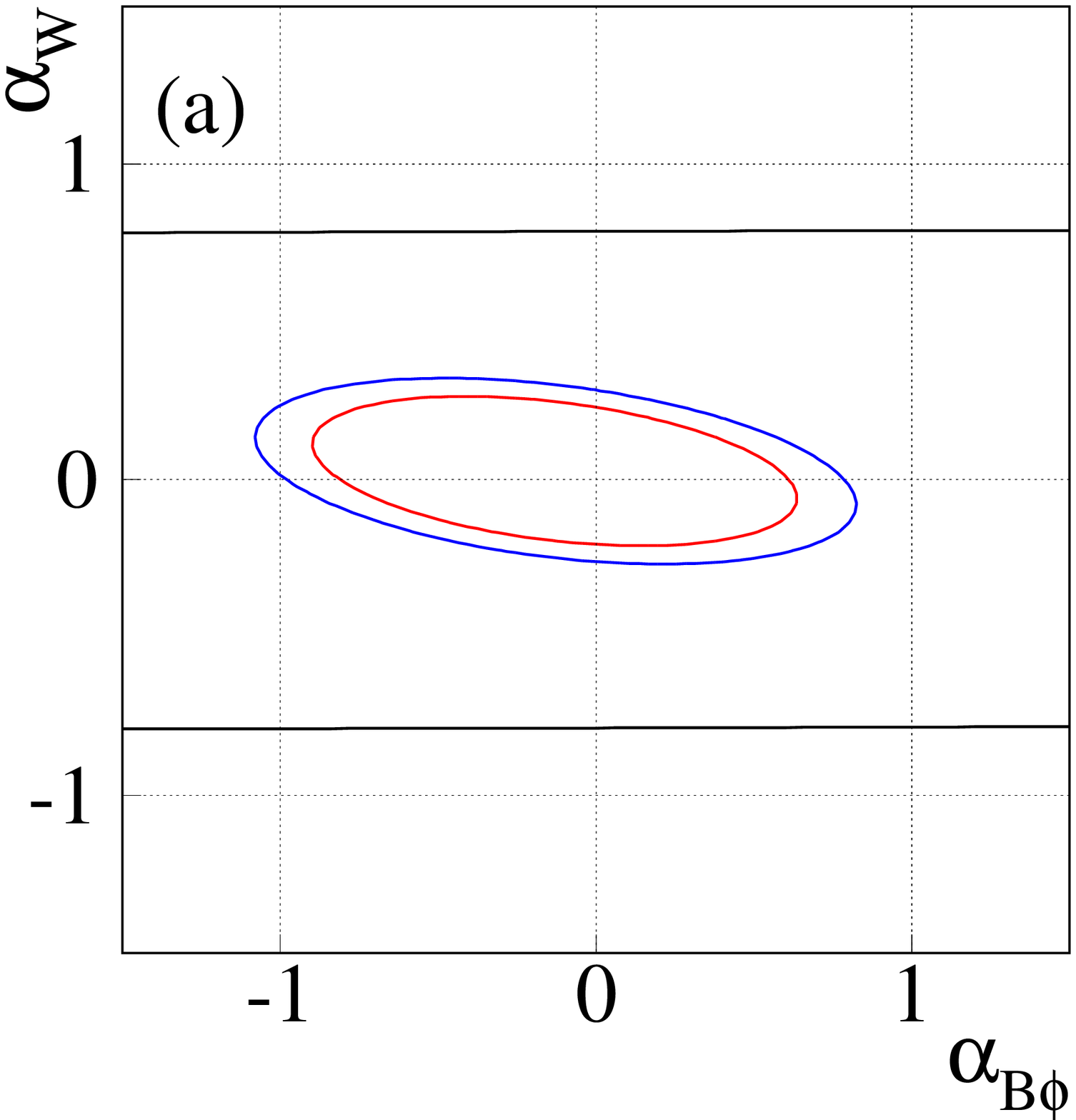,width=1.8in}
\epsfig{figure=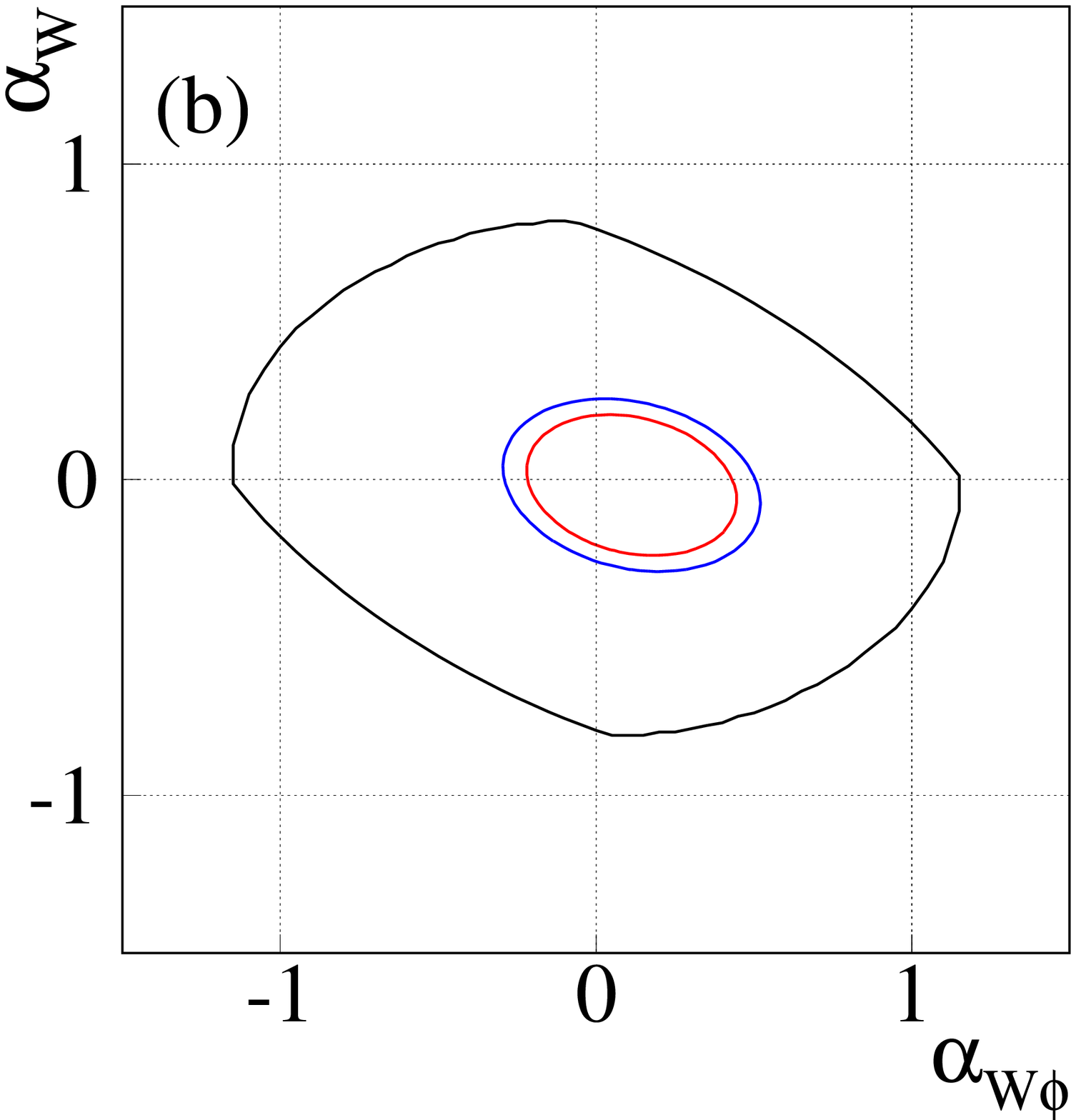,width=1.8in}}
}
 \caption{Contour limits on anomalous couplings
 from a simultaneous fit to the data sets of the $W\gamma$, $WW\rightarrow$
 dilepton, and $WW/WZ\rightarrow e\nu jj$ final states
 for $\Lambda = 1.5$ TeV:
 (a) $\alpha_{W}$ vs $\alpha_{B\phi}$ when $\alpha_{W\phi} = 0$; and
 (b) $\alpha_{W}$ vs $\alpha_{W\phi}$ when $\alpha_{B\phi} = 0$.
 The innermost and middle curves are
 $95 \%$ C.L. one- and two-degree of freedom exclusion contours, respectively.
 The outermost curve is the constraint from the unitarity
 condition.}
 \label{fig:lep_sets}
\end{figure}

In Table~\ref{table:combined_limits_alpha}, the $95 \%$ C.L. limits on
the $\alpha$ parameters from
a simultaneous fit to the three data sets are presented.
Limits on $\Delta g_1^Z$ are obtained,
from the limits on $\alpha_{W\phi}$ for
$\alpha_{B\phi}=\alpha_W=0$.
For comparison, limits from the OPAL collaboration~\cite{LEP} are also listed
in Table~\ref{table:combined_limits_alpha}. Limits from other LEP
collaborations are similar to those from OPAL.
Figure~\ref{fig:lep_sets}(a) shows the contour limits in the
$\alpha_W$-$\alpha_{B\phi}$ plane, when $\alpha_{W\phi} = 0$.
Figure~\ref{fig:lep_sets}(b) shows the contour limits in the
$\alpha_W$-$\alpha_{W\phi}$ plane, when $\alpha_{B\phi} = 0$.

In summary, limits on the anomalous $WW\gamma$ and $WWZ$ couplings are
obtained from a simultaneous fit to the data samples of three
gauge boson pair final states.
These limits are the tightest limits available on the anomalous $WW\gamma$ and
$WWZ$ couplings.

%
We thank the staffs at Fermilab and collaborating institutions for their
contributions to this work, and acknowledge support from the
Department of Energy and National Science Foundation (U.S.A.),
Commissariat  \` a L'Energie Atomique (France),
State Committee for Science and Technology and Ministry for Atomic
   Energy (Russia),
CAPES and CNPq (Brazil),
Departments of Atomic Energy and Science and Education (India),
Colciencias (Colombia),
CONACyT (Mexico),
Ministry of Education and KOSEF (Korea),
and CONICET and UBACyT (Argentina).

\begin{table}[htb]
\caption{Limits at 95\% C.L. from the $W\gamma$ analysis.}
\begin{tabular}{cc}
&$\Lambda=1.5$ TeV\\ \hline
$\lambda_{\gamma}$ ($\Delta\kappa_{\gamma}=0$)& $-0.31,~0.29$\\
$\Delta\kappa_{\gamma}$ ($\lambda_{\gamma}=0$)& $-0.93,~0.94$\\
\end{tabular}
\label{table:Wg_limits}
\end{table}

\begin{table}[htb]
\caption{Limits at 95\% C.L. from the $WW\rightarrow$ dilepton analysis.}
\begin{tabular}{cc}
&$\Lambda=1.5$ TeV\\ \hline
$\lambda_{\gamma} = \lambda_Z$ ($\Delta\kappa_{\gamma}=\Delta\kappa_Z=0$)&
 $-0.53,~0.56$\\
$\Delta\kappa_{\gamma} = \Delta\kappa_Z$ ($\lambda_{\gamma}=\lambda_Z=0$)&
 $-0.62,~0.77$\\ \hline
$\lambda_{\gamma}$(HISZ) ($\Delta\kappa_{\gamma}=0$)&
 $-0.53,~0.56$\\
$\Delta\kappa_{\gamma}$(HISZ) ($\lambda_{\gamma}=0$)&
 $-0.92,~1.20$\\
\end{tabular}
\label{table:dileptons_limits}
\end{table}

\begin{table}[htb]
\caption{Limits at 95\% C.L. from the $WW/WZ\rightarrow e\nu jj$ analysis.}
\begin{tabular}{ccc}
$\Lambda$&$1.5$ TeV&$2.0$ TeV\\ \hline
$\lambda_{\gamma} = \lambda_Z$ ($\Delta\kappa_{\gamma}=\Delta\kappa_Z=0$)&
 $-0.36,~0.39$& $-0.34,~0.36$\\
$\Delta\kappa_{\gamma} = \Delta\kappa_Z$ ($\lambda_{\gamma}=\lambda_Z=0$)&
 $-0.47,~0.63$& $-0.43,~0.59$\\ \hline
$\lambda_{\gamma}$(HISZ) ($\Delta\kappa_{\gamma}=0$)&
 $-0.36,~0.39$& $-0.34,~0.36$\\
$\Delta\kappa_{\gamma}$(HISZ) ($\lambda_{\gamma}=0$)&
 $-0.56,~0.85$& $-0.53,~0.78$\\ \hline
$\lambda_Z$(SM $WW\gamma$) ($\Delta\kappa_Z=\Delta g^Z_1=0$)&
 $-0.40,~0.43$& $-0.37,~0.40$\\
$\Delta\kappa_Z$(SM $WW\gamma$) ($\lambda_Z=\Delta g^Z_1=0$)&
 $-0.60,~0.79$& $-0.54,~0.72$\\
$\Delta g^Z_1$(SM $WW\gamma$) ($\lambda_Z=\Delta\kappa_Z=0$)&
 $-0.64,~0.89$& $-0.60,~0.81$\\ \hline
$\lambda_{\gamma}$(SM $WWZ$) ($\Delta\kappa_{\gamma}=0$)&
 $-1.21,~1.25$& --\\
$\Delta\kappa_{\gamma}$(SM $WWZ$) ($\lambda_{\gamma}=0$)&
 $-1.38,~1.70$& --\\
\end{tabular}
\label{table:enjj_limits}
\end{table}

\begin{table}[htb]
\caption{Limits at 95\% C.L. from a simultaneous fit to the $W\gamma$,
$WW\rightarrow$ dilepton and $WW/WZ\rightarrow e\nu jj$ data samples.
The four sets of limits apply the same assumptions as the four components
(a), (b), (c) and (d), respectively, of Fig.~\ref{fig:combined}.}
\begin{tabular}{ccc}
$\Lambda$&$1.5$ TeV&$2.0$ TeV\\ \hline
$\lambda_{\gamma} = \lambda_Z$ ($\Delta\kappa_{\gamma}=\Delta\kappa_Z=0$)&
 $-0.21,~0.21$& $-0.20,~0.20$\\
$\Delta\kappa_{\gamma} = \Delta\kappa_Z$ ($\lambda_{\gamma}=\lambda_Z=0$)&
 $-0.33,~0.46$& $-0.30,~0.43$\\ \hline
$\lambda_{\gamma}$(HISZ) ($\Delta\kappa_{\gamma}=0$)&
 $-0.21,~0.21$& $-0.20,~0.20$\\
$\Delta\kappa_{\gamma}$(HISZ) ($\lambda_{\gamma}=0$)&
 $-0.39,~0.61$& $-0.37,~0.56$\\ \hline
$\lambda_Z$(SM $WW\gamma$) ($\Delta\kappa_Z=\Delta g^Z_1=0$)&
 $-0.33,~0.37$& $-0.31,~0.34$\\
$\Delta\kappa_Z$(SM $WW\gamma$) ($\lambda_Z=\Delta g^Z_1=0$)&
 $-0.46,~0.64$& $-0.42,~0.59$\\
$\Delta g^Z_1$(SM $WW\gamma$) ($\lambda_Z=\Delta\kappa_Z=0$)&
 $-0.56,~0.86$& $-0.52,~0.78$\\ \hline
$\lambda_{\gamma}$(SM $WWZ$) ($\Delta\kappa_{\gamma}=0$)&
 $-0.27,~0.25$& $-0.26,~0.24$\\
$\Delta\kappa_{\gamma}$(SM $WWZ$) ($\lambda_{\gamma}=0$)&
 $-0.63,~0.75$& $-0.59,~0.72$\\
\end{tabular}
\label{table:combined_limits}
\end{table}

\begin{table}[htb]
\caption{Limits at 95\% C.L. on $\alpha$ parameters from a simultaneous fit
to the $W\gamma$, $WW\rightarrow$
dilepton and $WW/WZ\rightarrow e\nu jj$ data samples. Limits from the
OPAL collaboration are also listed for comparison.}
\begin{tabular}{ccc|c}
$\Lambda$&$1.5$ TeV&$2.0$ TeV&OPAL\\ \hline
$\alpha_{B\phi}$ ($\alpha_{W\phi}=\alpha_W=0$)&
 $-0.81,~0.61$& $-0.77,~0.58$& $-1.6,~2.7$\\
$\alpha_{W\phi}$ ($\alpha_{B\phi}=\alpha_W=0$)&
 $-0.24,~0.46$& $-0.22,~0.44$& $-0.55,~0.64$\\
$\alpha_W$ ($\alpha_{B\phi}=\alpha_{W\phi}=0$)&
 $-0.21,~0.21$& $-0.20,~0.20$& $-0.78,~1.19$\\
$\Delta g^Z_1$ ($\alpha_{B\phi}=\alpha_W=0$)&
 $-0.31,~0.60$& $-0.29,~0.57$& $-0.75,~0.77$\\
\end{tabular}
\label{table:combined_limits_alpha}
\end{table}


\begin{references}
\bibitem{CL} One- and two-degree of freedom exclusion contours are obtained
by cutting the log likelihood surface at 1.92 and 3.0 below the
maximum. These values correspond to the 95\% confidence levels for the cases
when only one free parameter is allowed (1.92) and two free parameters are
allowed (3.00) in the fit.
\end{references}
\end{document}